\documentclass[nohyper,12pt,letterpaper]{JHEP3}
\usepackage[dvips]{epsfig}
\usepackage{amsfonts,amssymb}

\newcommand{\be}{\begin{equation}}
\newcommand{\ee}{\end{equation}}
\newcommand{\ben}{\begin{displaymath}}
\newcommand{\een}{\end{displaymath}}
\newcommand{\bea}{\begin{eqnarray}}
\newcommand{\eea}{\end{eqnarray}}
\newcommand{\bean}{\begin{eqnarray*}}
\newcommand{\eean}{\end{eqnarray*}}

\newcommand{\ads}[1]{\mbox{${AdS}_{#1}$}}
\newcommand{\adss}[2]{\mbox{$AdS_{#1}\times {S}^{#2}$}}

\newcommand{\ket}[1]{\mbox{$| #1 \rangle$}}

\newcommand{\ie}{{\it i.e.}}

\newcommand{\tr}{\mbox{Tr}}

\newcommand{\beq}{\begin{equation}}
\newcommand{\eeq}{\end{equation}}
\newcommand{\beqr}{\begin{displaymath}}
\newcommand{\eeqr}{\end{displaymath}}
\newcommand{\beqa}{\begin{eqnarray}}
\newcommand{\eeqa}{\end{eqnarray}}
\newcommand{\beqar}{\begin{eqnarray*}}
\newcommand{\eeqar}{\end{eqnarray*}}
\newcommand{\cN}{{\cal N}}

\newcommand{\non}{\nonumber}

\newcommand{\half}{\ensuremath{\frac{1}{2}}}

\newcommand{\N}[1]{\ensuremath{\cN=#1}}

\title{\LARGE Spiky strings and single trace operators in gauge theories}

\author{Martin Kruczenski \\
        Department of Physics, Brandeis University \\
        Waltham, MA 02454.

E-mail: \email{martink@brandeis.edu}}

\abstract{ We consider single trace operators of the form 
$\mathcal{O}_{l_1\ldots l_n}=\tr D_+^{l_1} F \ldots D_+^{l_n} F$ 
which are common to all gauge theories. We argue that, when all $l_i$ are equal and
large, they have a dual description as strings with cusps, or spikes, one for each 
field $F$. In the case of $\mathcal{N}=4$ SYM, we compute the energy
as a function of angular momentum by finding the corresponding solutions in $AdS_5$ and compare
with a 1-loop calculation of the anomalous dimension. As in the case of two spikes (twist two operators), 
there is agreement in the functional form but not in the coupling constant dependence. After that,
we analyze the system in more detail and find an effective classical mechanics describing the motion
of the spikes. In the appropriate limit, it is the same (up to the coupling constant dependence) 
as the coherent state description of linear combinations of the operators $\mathcal{O}_{l_1\ldots l_n}$  
such that all $l_i$ are equal on average. This agreement provides a map between the operators in the boundary and 
the position of the spikes in the bulk. We further suggest that moving the spikes in other directions 
should describe operators with derivatives other than $D_+$ indicating that these ideas are quite generic
and should help in unraveling the string description of the large-N limit of gauge theories.
}


\preprint{\tt{BRX TH-555} \\
          \tt{hep-th/0410226}  }

\begin{document}

\section{Introduction}

 It has since long been suspected that four dimensional confining theories have a dual string description, at least
in the large-N limit~\cite{largeN}. The strings should emerge as the flux tubes of the confining force.
 However it was only relatively recently that a precise duality between a gauge theory and a string theory 
was established through the AdS/CFT correspondence~\cite{malda}. It turned out however that the theory in 
question (\N{4} SYM dual to $IIB$ strings on \adss{5}{5}) is not confining and therefore another explanation 
for the appearance of strings was required. In a recent paper, Berenstein, 
Maldacena and Nastase~\cite{Berenstein:2002jq} made the first steps in that direction by showing
that certain operators in the boundary theory corresponded to string excitation in the bulk. After
that, it was observed that such relation followed from a more general relation between that of semi-classical rotating 
strings in the bulk~\cite{GKP} and certain operators in the boundary. A lot of activity followed those papers. 
In particular many new rotating solutions were found\footnote{See 
the recent reviews~\cite{Tseytlin:2003ii,Tseytlin:2004xa,Tseytlin:2004cj} for a summary  
with a complete set of references.}. In a parallel development, Minahan and Zarembo~\cite{Minahan:2002ve} observed 
that the one-loop anomalous dimension of operators composed of scalars in \N{4} SYM theory follows from solving and 
integrable spin chain\footnote{In QCD the relation between spin chains and anomalous dimensions had already been noted 
in~\cite{Braun:1998id}. This relation is the one we are actually using in this paper.}. 
This allowed the authors of~\cite{Beisert:2003ea} to make a much more detailed comparison between particular 
string solutions and certain operators in the gauge theory. This result followed from a detailed
analysis of the dilatation operator which was discussed in \cite{Beisert:2003tq,Beisert:2003jj}, rotating string 
solutions \cite{Frolov:2003xy} and previous calculations \cite{Beisert:2003xu}. Also,
an important role in the comparison is taken by the study of the integrable structures of both sides 
of the correspondence \cite{Arutyunov:2003uj,Arutyunov:2003rg}. 

 It was later suggested \cite{Kruczenski:2003gt} that a classical sigma model action follows from 
considering the low energy excitations of the spin chain. It turned out that such action agrees with a 
particular limit of the sigma model action that describes the propagation of the strings in the bulk. This 
idea was extended to other sectors including fermions and open strings in 
\cite{Dimov:2004qv,Hernandez:2004uw,Ryang:2004tq,Dimov:2004xi,Stefanski:2004cw,Kruczenski:2004cn,
Ideguchi:2004wm,Ryang:2004pu,Hernandez:2004kr,Bellucci:2004qr,Susaki:2004tg}
and to two loops in \cite{Kruczenski:2004kw}. In \cite{Kazakov:2004qf,Kazakov:2004nh} the relation to 
the Bethe ansatz approach was clarified. Further, similar possibilities were found in certain subsectors
of QCD \cite{Ferretti:2004ba}. An interesting summary on the relation between previous work in QCD and
the more recent work can be found in \cite{Belitsky:2004cz}. A different but related approach to 
the relation between the string and Yang-Mills operators has been successfully pursued 
in \cite{Mikhailov:2004qf,Mikhailov:2004xw,Mikhailov:2004au} where a geometric description in 
terms of light like world-sheets was found. In that respect see also \cite{Gorsky:2003nq} and the discussion 
of supersymmetry in \cite{Mateos:2003de}.

 However, most of that work referred to a situation where the string was moving, at least partially on 
the sphere of \adss{5}{5}. Nevertheless, solutions rotating purely on \ads{5} are known, an example being 
the folded rotating string of \cite{GKP} which was conjectured to be dual to twist two operators in the gauge 
theory. This relation was 
confirmed in \cite{Kruczenski:2002fb,Makeenko:2002qe} where it was shown that similar results for the anomalous 
dimension are obtained by using Wilson loops with cusps (following previous work in QCD \cite{Korchemsky}). 
Twist two operators are single trace operators of the form
\beq
\mathcal{O} = \tr\,\Phi_1\, ({\stackrel{\leftrightarrow}{D}}_+)^l\ \Phi_2
\eeq
 where $D_+$ denotes the covariant derivative in a light-cone direction $x_+$ ($x_+ = x+t$ in Minkowski and 
$x_+=x_1+ix_2$ in Euclidean signature) and $\Phi_1$, $\Phi_2$ indicate generic adjoint fields in the gauge
theory. In QCD the anomalous dimension of these operators play an important role in the understanding of
deep inelastic scattering \cite{DIS}.
 
 In our case, we are interested in single trace operators because they are the relevant ones in the large-N
limit of gauge theories and we would like to find a string description for them. The most  
generic single trace operator is of the form (suppressing Lorentz indices)
\beq
\mathcal{O} = \tr D^{l_1} \Phi_1 \ldots D^{l_n} \Phi_n
\eeq
that is, has an arbitrary number of fields $\Phi_i$ instead of just two, and the derivatives can be in any direction. 
As an intermediate step to the full understanding of these operators, in this paper we concentrate on the simpler case
\beq
\mathcal{O} = \tr D_+^{l_1} \Phi_1 \ldots D_+^{l_n} \Phi_n
\label{ops}
\eeq
 namely when all derivatives are along the same direction $x_+$. Moreover, we also consider that the number 
of derivatives is large ($l_i\gg 1$). In this limit the anomalous dimension is dominated by the contribution
of the derivatives and we can ignore the exact nature of the fields $\Phi_i$ (in principle they do not even need 
to be elementary fields, they can also be composite fields of small conformal dimension).  
In the case of twist two operators, there is an associated string~\cite{GKP} (which is a closed string folded 
on itself to resemble a line segment) meaning that strings can be associated to such operators even when the 
number of fields is small. This 
appears different of what was discussed in \cite{Kruczenski:2003gt} where it was shown that a sigma model/string 
action follows from
taking a continuum limit when the number of operators is very large, each operator being associated to a
portion of the string. However it is not actually different if we think of associating each portion of the string to each
covariant derivative. In that case, the operators $\Phi_i$ should appear as distinguished points along the string.
 In the case of twist two operators there are two fields $\Phi_{1,2}$ and two distinguished points, namely
the two cusps at the end points of the segment along which the string is folded. 
 By analogy we expect that if we have $n$ operators, there should be $n$ distinguished points. One can, therefore, 
conjecture that operators of the form (\ref{ops}) have a dual description in terms of rotating strings with $n$ cusps
as already suggested in \cite{Belitsky:2003ys}. We test this idea by considering such strings rotating in \ads{5} and comparing their 
energy with a 1-loop calculation in the field theory in the limit of large angular momentum $J$. The leading term in the anomalous
dimension is $n/2$ times the result for the case of two cusps and behaves as $\ln J$. In the field theory the 
result is reproduced, up to the coupling constant dependence, by considering and operator with equal number 
of derivatives acting on each field. In particular, this again implies that the total number of derivatives, 
namely the angular momentum, is a multiple of $n$, the number of fields.

 At the next order in $J$ which is order one, such operator is not an eigenstate of the dilatation operator.
To better understand the system we map the operators into configurations of a spin chain with $n$ sites, one 
for each field, and with an infinite number of states per site, labeled by $l_i$ ($i=\ldots n$), the number of 
derivatives. We then consider a coherent state description derived in \cite{Stefanski:2004cw} and 
further studied in  \cite{Stefanski:2004cw,Bellucci:2004qr,Kazakov:2004qf,Belitsky:2003ys}
(following previous ideas in  \cite{Minahan:2002ve,Braun:1998id,Beisert:2003tq,Beisert:2003jj,Belitsky:2003ys}).
This description assumes that each site $i$ of the chain is in a coherent state parameterized by two 
variables $\rho_i,\theta_i$. 
The average number of derivatives is $\langle l_i\rangle=\sinh^2(\rho_i)$ which is consider to be constant. When $l_1\gg1$, the action 
of the dilatation operator on these states can be described in terms of a classical action for the remaining 
variables $\theta_i$.

 On the string side we consider the possibility that the spikes move, shifting their angular position and
find an effective classical mechanics that describes this motion. Up to the coupling constant dependence
it agrees with the one derived from the coherent state picture in the regime where the number of derivatives
at each site is large (or equivalently the energy of each spike is large).

We should emphasize once more that this fact does not imply an agreement between string and gauge theory since the
couplings are different on both sides. In spite of that, we believe that the comparison is still remarkable 
because it means that there is a string description of the operators even at 1-loop. The only point is that
such string propagates in an \ads{5} space of different radius than at strong coupling.

 The organization of this paper is as follows: in section \S\ref{flat space} we study the solutions 
in flat space and in section \S\ref{AdS} we extend them to AdS space, deriving also an 
effective description in terms of the dynamics of cusps. In section \S\ref{field theory} we compare
this description with the 1-loop field theory description of the dual operators and finally
give our conclusions in section \S\ref{conclu}.

\section{Spiky strings in flat space}
\label{flat space}

 As with the folded string, it is convenient to start by studying the solutions in flat space and then generalize them
to AdS. This is so because in flat space the solutions are particularly simple\footnote{In fact, in flat space, these 
solutions are known \cite{flatc} as pointed out to me by M. Gomez-Reino and J.J. Blanco-Pillado.} and therefore all their
properties can be easily derived. This simplicity is manifest in conformal gauge where they become just
a superposition of a left and a right moving wave:
\beqa
 x &=& \lambda\, \cos\left((n-1)\ \sigma_+\right) + \lambda\, (n-1)\, \cos\left(\sigma_-\right) \\
 y &=& \lambda\, \sin\left((n-1)\ \sigma_+\right) + \lambda\, (n-1)\, \sin\left(\sigma_-\right) \\
 t &=& 2\, \lambda\, (n-1)\, \tau = \lambda\, (n-1)\,(\sigma_++\sigma_-)
\label{flatsol}
\eeqa
 where $\sigma_+ = \tau+\sigma$, $\sigma_-=\tau-\sigma$, $n$ is an integer and $\lambda$ is constant that
determines the size of the string. Here, ($\tau,\sigma$) parameterize the world-sheet of a string moving 
in a Minkowski space of metric:
\beq
 ds^2 = -dt^2 + dx^2+dy^2
\eeq
 The solutions are periodic in $\sigma$ with period $2\pi$ and satisfy the equations of motion in conformal gauge:
\beq
 (\partial_\tau^2 - \partial_\sigma^2) X^{\mu} = 0 , \ \ \ X^{\mu}=(t,x,y)
\eeq
as well as the constraints
\beq
 (\partial_+ X)^2 = (\partial_-X)^2 =0  
\eeq
 Quantum mechanically the state has $n_R = \lambda^2 (n-1)^2$ right moving excitations of wave number $k_R=1$ 
and $n_L = \lambda^2 (n-1)$ left moving excitations with wave number $k_L=n-1$ (satisfying the level matching 
condition $n_Rk_R =n_L k_L$). 

 All excitations carry one unit of angular momentum and therefore the total angular momentum and energy are given by 
\beq
J = n_L+n_R = \lambda^2 n (n-1) , \ \ E=\sqrt{2(n_Lk_L+n_Rk_R)}= 2\lambda (n-1), \ \ E = 2\sqrt{\frac{n-1}{n}J}
\eeq
which agrees with a classical computation. For $n=2$ we recover the standard Regge 
trajectory $E=\sqrt{2J}$ and for $n>2$ we get a Regge trajectory of modified slope. 
The variation in slope is not large since for $n\rightarrow \infty$ we get $E=2\sqrt{J}$ 
(however, this limit is singular because the string has infinite energy). 
 
At fixed time, the shape of the string for different values of $n$ takes the form  
depicted in figs.(\ref{fig:3flat}) and (\ref{fig:10flat}). It can be seen that the string has $n$ 
spikes\footnote{It might be interesting to consider also T-dual configurations which have $(n-2)$ spikes
pointing inwards.} or cusps and,
 analyzing the time dependence, that it rotates rigidly in such a way that the end points of the spikes
move at the speed of light. Notice that this does not necessarily mean that a point of fixed $\sigma$ moves in a circle
since the string can slide on itself. In the appendix we introduce other world-sheet coordinates where $\sigma$ 
parameterizes the shape of the string and translations in $\tau$ correspond to rigid rotations. It is clear however 
that in such a gauge $\dot{X}.X'\neq 0$ and therefore it is not the conformal gauge used in this section. 

 To find the position of the cusps we can use the fact they appear whenever $X'^2=0$ which happens if
\beq
 (n-1)\sigma_+-\sigma_- = 2\pi m, \ \ \Rightarrow \ \ \sigma=2\pi \frac{m}{n} -\frac{n-2}{n} \tau\ 
\ \ \mbox{with} \ \ m=0\ldots n-1\ .
\eeq
Therefore, there are $n$ cusps which, for $\tau=0$, are at $\sigma=2\pi \frac{m}{n}$, $m=0\ldots n-1$. In space time
they are located at $r=r_c=n\lambda$ and $\theta=2\frac{n-1}{n} \tau - 2\pi \frac{m}{n}$  where $(r,\theta)$ are
standard polar coordinates in the plane $(x,y)$.  The velocity at which the spikes move is given by
\beq
 v = r_c \frac{d\theta}{dt} = r_c  \frac{d\theta}{d\tau}\frac{d\tau}{dt}= 1
\eeq
namely, the speed of light. 

As a final point, to which we will come back later, note that $J$ can be written as $J=n_Ln$ and therefore
is a multiple of $n$. The same result can be argued using the classical solution\footnote{I am grateful to L. Urba
for providing this argument.}: since the shape of the string is invariant under rotations of $\frac{2\pi}{n}$, if
we quantize it as a rigid body, the only allowed values of $J$ are multiples of $n$ in such a way that the wave
function has the same symmetry under $\frac{2\pi}{n}$ rotations. More precisely, if $\ket{\theta}$ denotes a state
where the configuration is rotated by an angle $\theta$, a state of angular momentum $J$ is
\beq
\ket{J} = \int_0^{2\pi} e^{iJ\theta}\ket{\theta} \, d\theta
\eeq
If $\ket{\theta} = \ket{\theta+\frac{2\pi}{n}}$, the integral vanishes unless $J$ is a multiple of $n$. 

The argument based on the shape of the classical solution becomes important in the \ads{5} case where 
we cannot quantize the string exactly. 

In the next section we generalize these solutions to AdS space. Before doing that however let us comment
in one counterintuitive feature of the flat space solutions. In the folded rotating string one  
argues that the tension of the string provides the centripetal force necessary for circular motion. 
The same idea applies for the spikes here but, on the other hand, the bottom of the valleys
should be pulled outwards by the rest of the string, opposite to the centripetal acceleration. 
 The first point to explain what happens is that there is no centripetal acceleration since the 
valley is instantaneously at rest as a simple calculation shows and as also follows from the fact that,
at the bottom of the valley, any angular motion is parallel to the string and therefore unphysical. 
 Nevertheless it is still true that the rest of the string pulls this point outwards in the radial direction.
 Again a simple computation reveals that the point moves away from the origin in such a way that it
still stays on the string profile. Namely, the string rotates and therefore an instant later, at the same 
angular position, the radial position of the string increased exactly to match the outward motion of the 
portion of string at the valley.

\FIGURE{\epsfig{file=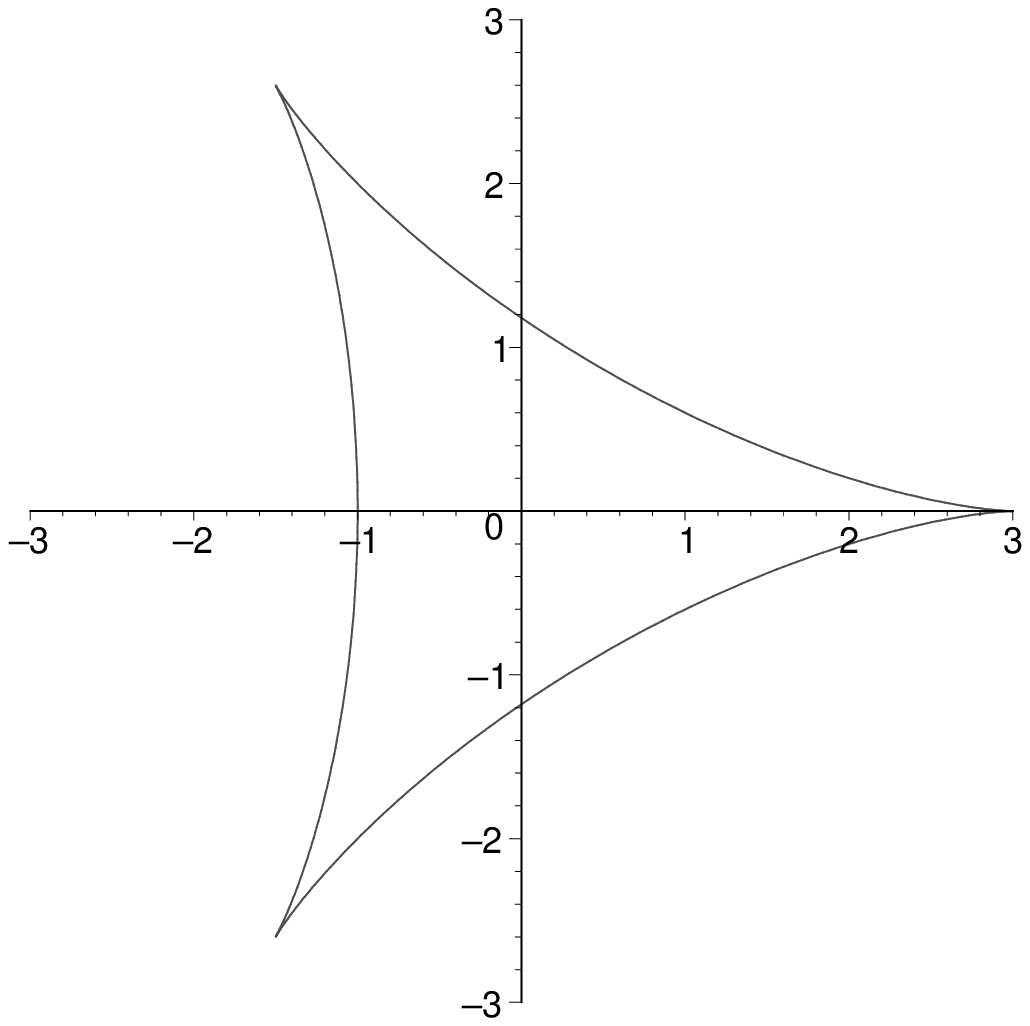, height=10cm}
\caption{Rotating string with 3 spikes in flat space. We plot $(x,y)$ in eq.(\ref{flatsol}) 
parametrically as a function of $\sigma=0\rightarrow2\pi$ for $n=3$, $\lambda=1$ and $\tau=0$.
 }
\label{fig:3flat} }
\FIGURE{\epsfig{file=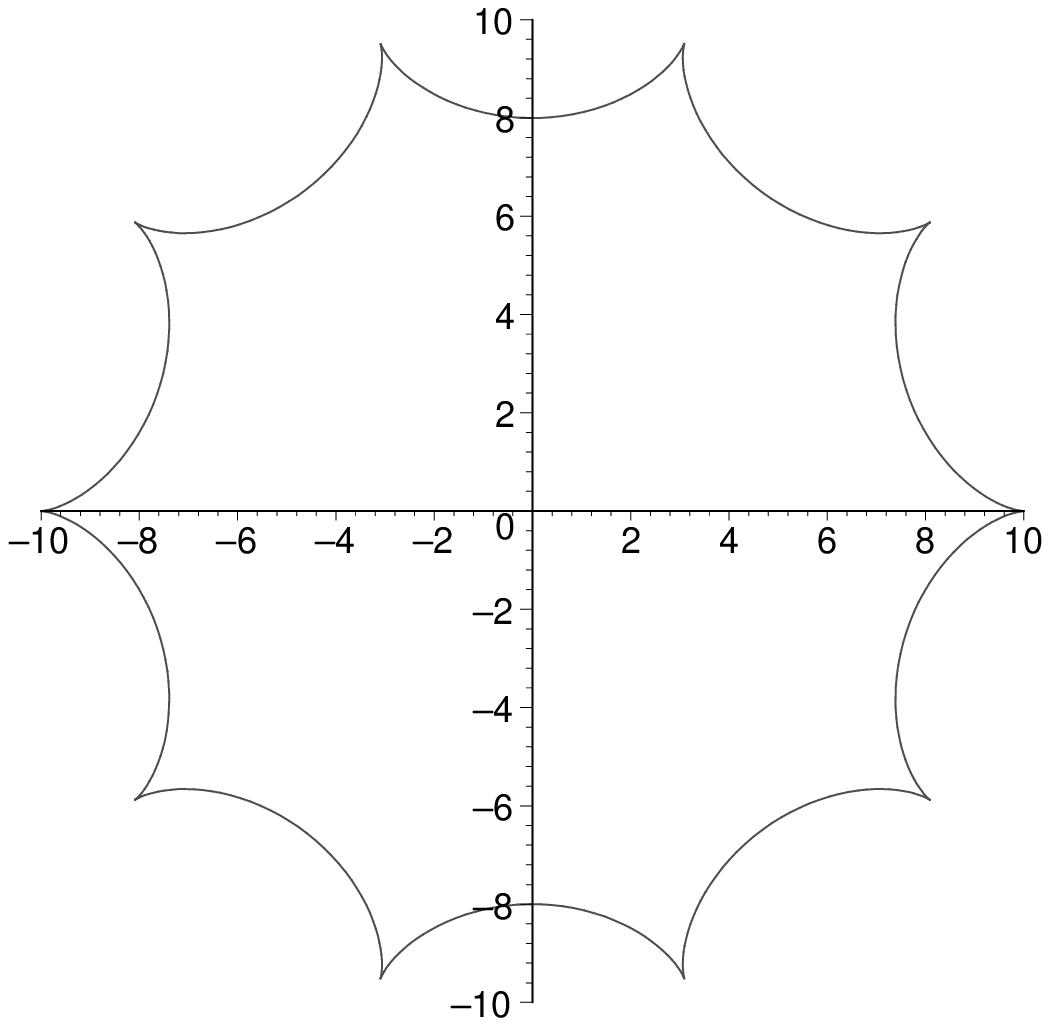, height=10cm}
\caption{Rotating string with 10 spikes in flat space. We plot $(x,y)$ in eq.(\ref{flatsol}) 
parametrically as a function of $\sigma=0\rightarrow2\pi$ for $n=10$, $\lambda=1$ and $\tau=0$.}
\label{fig:10flat} }
\FIGURE{\epsfig{file=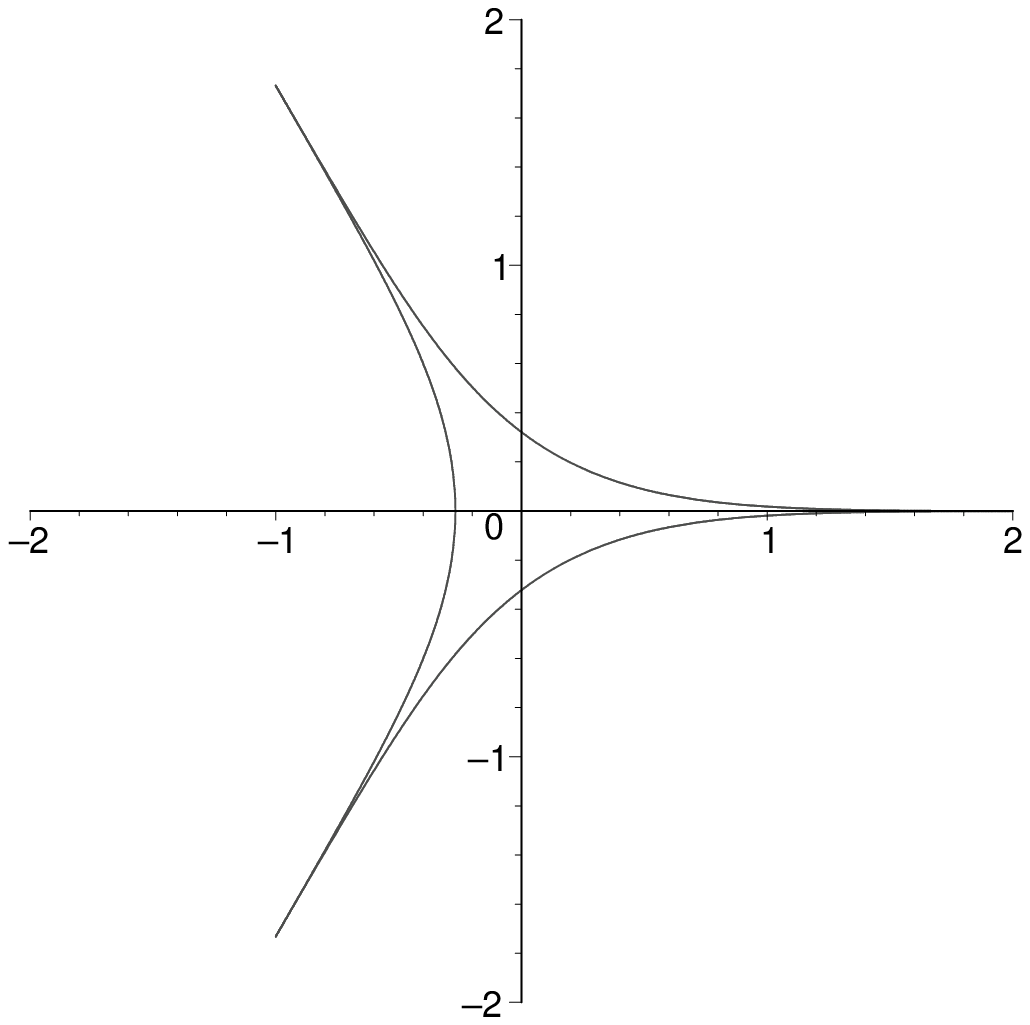, height=10cm}
\caption{Rotating string with 3 spikes in AdS space. In the figure, $(\rho,\theta)$ in eq.(\ref{AdSmetric}) 
are standard polar coordinates and we use the ansatz (\ref{soc}) together with (\ref{alpha}) (and $\tau=0$) 
to plot one half of a spike and then rotate and reflect to plot the others. The same result can be obtained by using 
(\ref{noglue}) which does not require this gluing procedure. In both cases the parameters are 
$\rho_0=0.2688735$, $\rho_1=2$. }
\label{fig:3ads} }
\FIGURE{\epsfig{file=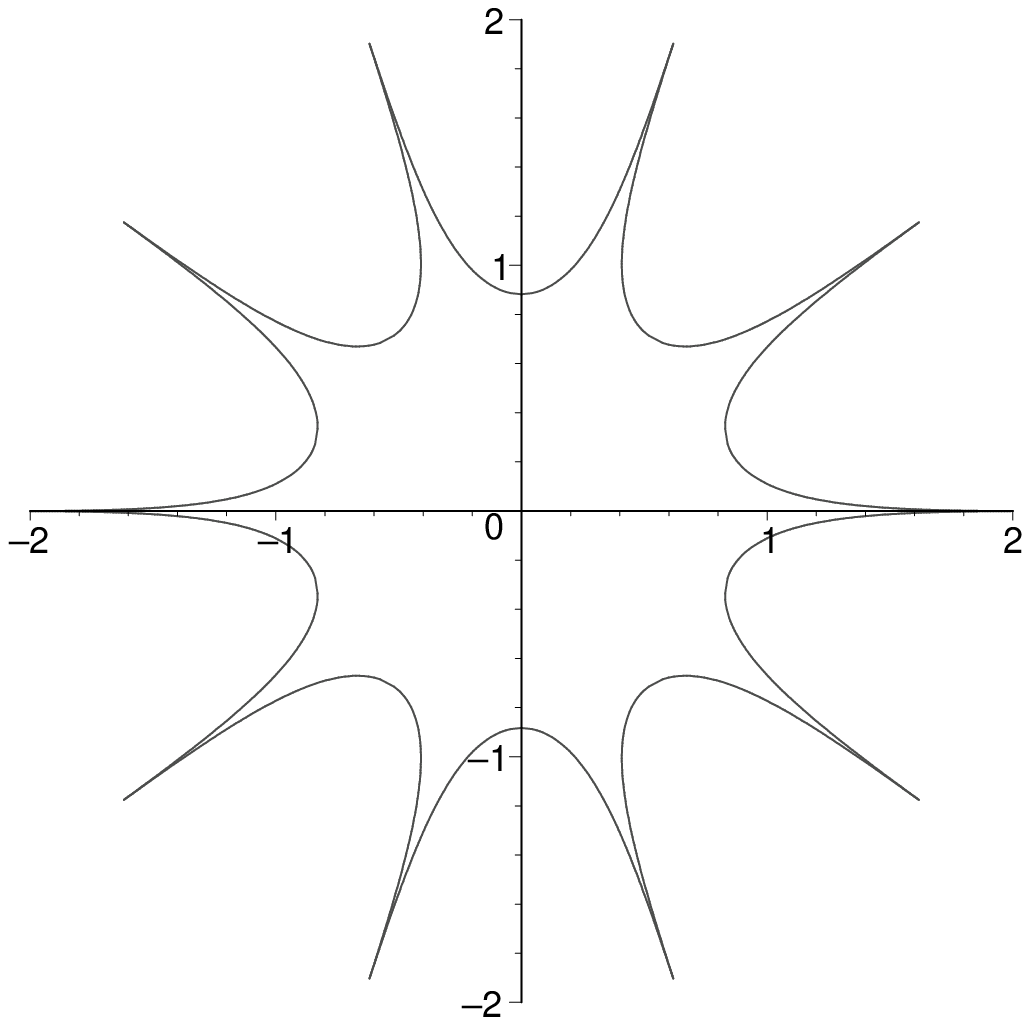, height=10cm}
\caption{Rotating string with 10 spikes in AdS space obtained in the same way as fig.\ref{fig:3ads} but with
$\rho_0=0.88266273$, $\rho_1=2$.}
\label{fig:10ads} }

\section{Spiky strings in AdS}
\label{AdS}

 For the case of string rotating in AdS we were not able to find such simple expressions as for the
flat space case. We resort to using the Nambu-Goto action which leads to more complicated expressions
but whose properties are still easily understood. Consider a string moving in a space parameterized by
$(t,\rho,\theta)$ with metric 
\beq
 ds^2 = -\cosh^2\!\rho\, dt^2 + d\rho^2 + \sinh^2\!\rho\, d\theta^2
\label{AdSmetric}
\eeq
which is the metric of \ads{3} which can be considered as embedded in \ads{5} for the purposes of the
AdS/CFT correspondence. We choose world-sheet coordinates in such a way that 
\beq
 t= \tau , \ \ \theta = \omega\tau + \sigma,
\label{soc}
\eeq
and make the ansatz that the string is rigidly rotating which implies $\rho=\rho(\sigma)$, 
namely $\partial_\tau\rho=0$. 
 The Nambu-Goto action is 
\beq
 S = - \frac{\sqrt{\lambda}}{2\pi} \int \sqrt{-\dot{X}^2 {X'}^2 + (\dot{X}X')^2}
\label{NGaction}
\eeq
where $\lambda$ is the 't Hooft coupling in the field theory and the scalar products are taken 
with the metric (\ref{AdSmetric}). Notice that we set the radius $R$ of \ads{5} to one by writing explicitly the
't Hooft coupling whenever $R^2/\alpha'$ should have appeared. This makes all quantities adimensional.
The ansatz we use implies that
\beq
\dot{X}^2  = -\cosh^2\rho + \omega^2 \sinh^2\rho, \ \ {X'}^2 = {\rho'}^2 + \sinh^2\rho , \ \ \dot{X}X' = \omega \sinh^2\rho
\eeq
and 
\beq
 \sqrt{-\dot{X}^2 {X'}^2 + (\dot{X}X')^2} = \sqrt{{\rho'}^2(\cosh^2\rho-\omega^2\sinh^2\rho)+\sinh^2\rho\cosh^2\rho}
\eeq
 We can further check that the equations of motion for $t$ and $\theta$, following from the action (\ref{NGaction}), are
satisfied if
\beq
\rho' = \half\frac{\sinh2\rho}{\sinh2\rho_0}\frac{\sqrt{\sinh^22\rho-\sinh^22\rho_0}}{\sqrt{\cosh^2\rho-\omega^2\sinh^2\rho}}
\label{rhoeq}
\eeq
where $\rho_0$ is an integration constant. Furthermore we can check that, if we assume (\ref{rhoeq}), 
the equation of motion for $\rho$ is also satisfied. From the expression for $\rho'$ we see that $\rho$ varies
from a minimum value $\rho_0$ to a maximum value $\rho_1=\mbox{arccoth}\,\omega$. At $\rho=\rho_1$, $\rho'$ diverges
indicating the presence of a spike and at $\rho=\rho_0$ vanishes, indicating the bottom of the valley between spikes. To get
a solution with $n$ spikes we should glue $2n$ of the arc segments we get here. For that we need to choose $\rho_1$ and 
$\rho_0$ in such a way that the angle between cusp and valley is $\frac{2\pi}{2n}$. This fixes one constant. The other,
for example, can be fixed by giving the total angular momentum $J$. In that way we can determine the energy $E(n,J)$.

\subsection{Exact solution}

If we take equation (\ref{rhoeq}) for $\rho$ and do the following change of variables: 
\beq
u = \cosh2\rho, \ \ \ u_0=\cos2\rho_0, \ \ \ u_1=\cosh2\rho_1 .
\label{udef}
\eeq
 the resulting integral can be computed in terms of the elliptic function $\Pi$ defined as \cite{Grad}
\beq
 \Pi(\alpha,n,p) = \int_0^{\alpha} \frac{d\beta}{\sqrt{1-p^2\,\sin^2\!\!\beta}\ (1-n\,\sin^2\!\!\beta)} , \ \ \ \ 
\alpha\le\frac{\pi}{2}
\label{EllipticPi}
\eeq
 The result is
\beqa
\sigma &=& \frac{\sinh2\rho_0}{\sqrt{2}\sqrt{u_0+u_1}\sinh\rho_1}
  \left\{\Pi(\alpha,\frac{u_1-u_0}{u_1-1},p)-\Pi(\alpha,\frac{u_1-u_0}{u_1+1},p)  \right\}
\label{alpha}
\eeqa
where
\beqa
 p &=& \sqrt{\frac{u_1-u_0}{u_1+u_0}} \label{pdef}\\
 \sin\alpha &=&  \sqrt{\frac{u_1-u}{u_1-u_0}} \label{alphadef}
\eeqa
 The resulting shape, for constant time and two different values of $n$, is depicted in figs.(\ref{fig:3ads}) and (\ref{fig:10ads}). 
The angle difference between the spike and the valley as well as the angular momentum and energy carried by the string
can be computed as
\beqa
\Delta \theta &=&  2 \int_{\rho_0}^{\rho_1}\frac{\sinh2\rho_0}{\sinh2\rho}
 \frac{\sqrt{\cosh^2\rho-\omega^2\sinh^2\rho}}{\sqrt{\sinh^22\rho-\sinh^22\rho_0}}\, d\rho \\\non \\
J &=& \frac{2n}{2\pi}\frac{\omega}{2}  \sqrt{\lambda} \int_{\rho_0}^{\rho_1}\frac{\sinh\rho}{\cosh\rho}  
      \frac{\sqrt{\sinh^22\rho-\sinh^22\rho_0}}{\sqrt{\cosh^2\rho-\omega^2\sinh^2\rho}}\, d\rho \label{Jint}\\\non \\
E-\omega J &=& \frac{2n}{2\pi}\half \sqrt{\lambda}\int_{\rho_0}^{\rho_1}
  \sinh2\rho\frac{\sqrt{\cosh^2\rho-\omega^2\sinh^2\rho}}{\sqrt{\sinh^22\rho-\sinh^22\rho_0}}\, du 
\eeqa
As discussed before, $\lambda$ is the 't Hooft coupling, and $n$ the number of spikes which 
determines $\Delta\theta=\frac{2\pi}{2n}$. Note also that
$E$ and $J$ are multiplied by $2n$ to obtain the total energy and angular momentum. 
Again, the change of variables (\ref{udef}) proves useful reducing the computation to 
elliptic integrals:
\beqa
\rule{0cm}{0.2cm}\non\\
 \Delta \theta &=& \frac{\sinh2\!\rho_0}{\sqrt{2}\sinh\rho_1}\frac{1}{\sqrt{u_1-u_0}}
 \left\{\Pi(\frac{\pi}{2},\frac{u_1-u_0}{u_1-1},p)-\Pi(\frac{\pi}{2},\frac{u_1-u_0}{u_1+1},p) \right\} 
 \\\rule{0cm}{0.2cm}\non\\
 J &=& \frac{2n}{2\pi}\ \frac{\sqrt{\lambda}}{\sqrt{2}}\ \frac{\cosh\rho_1}{\sqrt{u_1+u_0}}\ \bigg\{
    -(1+u_0) \mathbf{K}(p) + (u_1+u_0) \mathbf{E}(p) -  \non\\
 && \ \ \ \ \ \ \ \  \frac{u^2_0-1}{u_1+1} \Pi(\frac{\pi}{2},\frac{u_1-u_0}{u_1+1},p) \bigg\}
\\\rule{0cm}{0.2cm}\non\\
E-\omega J &=& \sqrt{\lambda}\ \frac{2n}{2\pi}\ \frac{\sqrt{u_1+u_0}}{\sqrt{2}\sinh\rho_1} 
  \ \left( \mathbf{K}(p)-\mathbf{E}(p)\right)
\\\rule{0cm}{0.2cm}\non
\eeqa
with $p$ as in (\ref{pdef}). The functions $\mathbf{K(p)}$ and $\mathbf{E}(p)$ are defined by~\cite{Grad}:
\beq
\mathbf{K}(p) = \int_0^{\frac{\pi}{2}} \frac{d\alpha}{\sqrt{1-p^2\sin^2\alpha}}, \ \ \ 
\mathbf{E}(p) = \int_0^{\frac{\pi}{2}} \sqrt{1-p^2\sin^2\alpha}\,d\alpha
\eeq
Although it is interesting that the result can be expressed in terms of well-studied functions the actual
expressions are not very illuminating and will not be used in the rest of the paper. For comparison with
the field theory we consider now the limit in which the string is moving in such a way that $\rho_1\gg1$
which, as we will see, is the large angular momentum limit. Before doing that however, we would like to 
clarify the process of gluing several segments to form the string. In the previous section that was not
necessary and here actually it is not necessary either. It rather appears because of the choice of
world-sheet coordinates. We can define a new $\tilde{\sigma}$ such that
\beqa
\theta -\omega\tau = \sigma &=& \frac{\sinh2\rho_0}{\sinh\rho_1\sqrt{2(u_1+u_0)}} \left\{
\int_0^{\tilde{\sigma}} \frac{d\alpha}{\sqrt{1-p^2\sin^2\alpha}\ [1-\frac{u_1-u_0}{u_1-1}\sin^2\alpha]} -
\right. \non \\
&& \left.\int_0^{\tilde{\sigma}} \frac{d\alpha}{\sqrt{1-p^2\sin^2\alpha}\ [1-\frac{u_1-u_0}{u_1+1}\sin^2\alpha]} 
   \right\} \\ \rule{0cm}{0.4cm}\non\\
\sinh^2\rho &=& \sinh^2\rho_1\cos^2\tilde{\sigma}+\sinh^2\rho_0\sin^2\tilde{\sigma} 
\label{noglue}
\eeqa
The integrals are of the form (\ref{EllipticPi}) and then at first sight it seems that $\tilde{\sigma}$ is
the same as $\alpha$ in (\ref{alpha}). However care should be taken since this is valid only for the first half 
spike (at least according to the standard definition of the elliptic integral which is valid 
for $\alpha\le\frac{\pi}{2}$)\footnote{This point is just a problem of definitions but should be taken into account
before using formulas which assume some particular 
extension of $\Pi(\alpha,n,p)$  for $\alpha\ge\frac{\pi}{2}$. }. As defined here, 
$\sigma$ is a monotonically increasing function of 
$\tilde{\sigma}$, namely $\frac{d\sigma}{d\tilde{\sigma}}\ge 0$ with $\frac{d\sigma}{d\tilde{\sigma}}= 0$ precisely 
at the cusps since $\frac{d\rho}{d\sigma}=\frac{d\rho}{d\tilde{\sigma}}\frac{d\tilde{\sigma}}{d\sigma}$ diverges
if $\frac{d\sigma}{d\tilde{\sigma}}=0$. In fact, that is the only indication of the presence of cusps 
since now $\theta$ and $\rho$ are regular functions of $\tilde{\sigma}$ and no gluing process is needed. 
Of course, what we still need to do is to adjust $\rho_0$ and $\rho_1$ to get the right periodicity 
for $\theta(\tilde{\sigma})$ and $\rho(\tilde{\sigma})$.

\subsection{Asymptotic expansion}

To compare with the field theory we take a large angular momentum limit corresponding to
$\rho_1\gg1$. In this limit, the energy $E$ turns out to be equal, at leading order, to the 
angular momentum $J$ and the difference $E-J$, which we compare to the field theory, is proportional 
to $\sqrt{\lambda}(\rho_1-\rho_0)$. In view of this, we require, in addition, that $\sqrt{\lambda}(\rho_1-\rho_0)\gg1$ 
which means that the spike carries a large energy and can be described semi-classically\footnote{The 
limit $\rho_1-\rho_0\rightarrow 0$ presumably describes a regime where the contribution of the derivatives 
to the anomalous dimension is not dominant but it is not clear to us what operators describe that situation. }.

 In this limit, $\omega=\coth\rho_1\rightarrow 1$, and a good approximation is obtained by replacing 
$\omega=1$ in $\Delta\theta$ and $E-\omega J$. This results in 
\beqa
E-J \simeq E-\omega J &\simeq&\frac{2n}{2\pi}\sqrt{\lambda}\int_{\rho_0}^{\rho_1}\!\!\!\frac{\sinh2\rho}{\sqrt{\sinh^22\rho-\sinh^22\rho_0}}\,d\rho
  = \frac{2n}{4\pi}\sqrt{\lambda}\, \mbox{arccosh}\frac{u_1}{u_0} \\\rule{0cm}{0.2cm}\non\\
\Delta\theta &\simeq& 
   2\int_{\rho_0}^{\rho_1}\frac{\sinh2\rho_0}{\sinh2\rho}\frac{d\rho}{\sqrt{\sinh^22\rho-\sinh^22\rho_0}} 
\\\rule{0cm}{0.2cm}\non \\ 
  &=& \frac{\pi}{4}-\half\arcsin\left(\frac{1+u_1^2-2u_1^2/u_0^2}{u_1^2-1}\right) \label{DTl}
\eeqa
where, as in (\ref{udef}) we defined $u_1=\cosh2\rho_1$ and $u_0=\cosh2\rho_0$. From here we obtain the energy
of each half-spike as
\beqa
\frac{E-J}{2n} &\simeq& \frac{\sqrt{\lambda}}{8\pi}
                    \mbox{arccosh}\left(1+2\sinh^22\rho_1\sin^2\Delta\theta\right) \label{En1}\\
               &\simeq& \frac{\sqrt{\lambda}}{8\pi}\left\{4\rho_1+\ln(\sin^2\Delta\theta)\right\} \label{En2}
\eeqa
where the last approximation is done using $\rho_1\gg 1$ and assumes that $\Delta\theta$ is not as small as to 
make $e^{2\rho_1}\sin\Delta\theta \gg 1$ invalid. This means that we cannot put an extremely large number of operators
$n\sim e^{2\rho_1}$. As shown below, this is equivalent to the assumption $\rho_1-\rho_0\gg1$.
The result (\ref{En1}) can also be derived easily in a more illuminating way. Indeed, the 
approximation $\omega\simeq 1$ can be done directly in the Nambu-Goto action resulting in 
\beqa
S_{NG} &=& -\frac{\sqrt{\lambda}}{2\pi}\int\sqrt{\rho'^2(\cosh^2\rho-\omega^2\sinh^2\rho)+\sinh^2\rho\cosh^2\rho} \\
       &\simeq& -\frac{\sqrt{\lambda}}{4\pi}\int \sqrt{(2\rho')^2+\sinh^22\rho}
\eeqa
This means that, in this limit, the action is simply the one for geodesics in the metric
\beq
ds^2 = (d\tilde{\rho})^2 + \sinh^2\!\tilde{\rho}\, d\theta^2
\label{effmet}
\eeq
where $\tilde{\rho}=2\rho$, we parameterize the geodesic with $\theta=\sigma$, and solve for $\rho(\sigma)$.
 The energy is therefore
\beq
\frac{E-J}{2n} = \frac{\sqrt{\lambda}}{8\pi} L(\rho_1,\Delta\theta)
\eeq
where $L(\rho_1,\Delta\theta)$ is the geodesic distance between two points at the same value of $\tilde{\rho}=2\rho_1$
and separated by an angle $2\Delta\theta$  (recalling that $\Delta\theta$ is the angle between a peak and a valley,
and therefore half the angle between peaks). The geodesic length in the hyperboloid (\ref{effmet}) can be easily 
computed and agrees with (\ref{En1}).

To compute $J$, we notice that the largest contribution comes from $\rho\sim\rho_1$ where the denominator
inside the integral (\ref{Jint}) is small. Again, assuming $\rho_1-\rho_0\gg1$, we can make the 
approximation that $\rho\sim\rho_1\gg\rho_0$ leading to
\beqa
J &\simeq& \frac{2n}{2\pi}\sqrt{\lambda}\int_{\rho_0}^{\rho_1}\!\!\!
            \frac{\sinh^2\rho}{\sqrt{\cosh^2\rho-\omega^2\sinh^2\rho}}\,d\rho 
           \simeq\frac{2n}{2\pi}\frac{\sqrt{\lambda}}{4}e^{\rho_1}\int_{\rho_0}^{\rho_1}\!\!\!
                \frac{e^{2\rho}}{\sqrt{e^{2\rho_1}-e^{2\rho}}}\,d\rho \non\\
  &\simeq& \frac{2n}{2\pi}\frac{\sqrt{\lambda}}{4}e^{2\rho_1} \label{Jl}
\eeqa

Let us now consider the limit in which $\rho_1\rightarrow \infty$ keeping $\rho_0$ fixed and large. In this limit
(\ref{DTl}) implies that
\beq
\Delta\theta \simeq \frac{1}{u_0} \simeq 2\  e^{-2\rho_0}
\eeq
and (\ref{Jl}) implies that $J\rightarrow\infty$. Since the number of spikes $n$ is equal to $n=\pi/\Delta\theta$, 
the limit $\rho_1\rightarrow \infty$ keeping $\rho_0$ fixed and large is the limit of large angular momentum
keeping the number of spikes fixed. In this limit we get, from (\ref{En2}) and for large $J$:
\beq
E \simeq J+\sqrt{\lambda}\ \frac{n}{2\pi}\, \ln\left(\frac{4\pi}{n}\frac{J}{\sqrt{\lambda}}\right), 
\ \ \ \ \ \ \ \ (J\rightarrow\infty)
\label{enres}
\eeq
which, for $n=2$ agrees with the result obtained in \cite{GKP} and in \cite{Kruczenski:2002fb,Makeenko:2002qe} 
using a different method. 

\subsection{Fluctuations}

 Certain fluctuations around the solution correspond to a change in the positions of the tip of the spikes along 
the circle $\rho=\rho_1$. The potential energy of interaction between spikes is given by the energy of
the string hanging between them as computed in (\ref{En2}), where we should now keep the (subleading)
term proportional to $\ln(\sin^2\Delta\theta)$ which contains the dependence on their relative position.
 A kinetic energy appears when the spikes move with respect to each other. 
Since the tip of the spike moves at the speed of light this can only happen if a spike moves up or down
in $\rho$. For example if it goes up the angular velocity decreases but at the same time there is a cost
in energy since the spike is longer. More precisely, if the spike shift upward to a position $\rho_1+\delta\rho$
the energy cost is 
\beq
 \delta (E - J) = 2\ \frac{\sqrt{\lambda}}{2\pi}\ \delta \rho
\eeq
since, at constant $\Delta \theta$, from (\ref{En2}) we get 
$E-J\simeq 2\, \frac{\sqrt{\lambda}}{2\pi} \rho_1$ (the extra factor of $2$ is because (\ref{En2}) corresponds
to half a spike). At the same time the variation in angular velocity is 
\beq
\delta \omega = \frac{d}{d\rho_1} \frac{\cosh\rho_1}{\sinh\rho_1}\  \delta\rho = - \frac{1}{\sinh^2\rho_1}\ \delta\rho
\eeq
From here we obtain
\beq
\delta (E-J) = -\frac{\sqrt{\lambda}}{2\pi}\ 2\sinh^2\!\rho_1\,\delta\omega 
 = \frac{\sqrt{\lambda}}{2\pi}\ (\cosh2\rho_1-1)\ \dot{\theta}
\eeq
where we call $\theta$ the displacement of the spike with respect to a reference position moving at constant 
angular velocity. $\delta (E-J)$ acts as the kinetic energy. Putting all together we expect the dynamics of
the spikes to be described by an action
\beq
 S = \frac{\sqrt{\lambda}}{2\pi}\int\!\! d\tau \sum_j (\cosh2\rho_1-1)\ \dot{\theta_j} -  
   \frac{\sqrt{\lambda}}{8\pi} \int\!\!d\tau\sum_j \left\{4\rho_1+\ln(\sin^2(\frac{\theta_{j+1}-\theta_j}{2}))\right\}
\eeq
where $\theta_j$ denotes the angular position of each spike assumed to move around the circle $\rho=\rho_1$.
In the following section we compare this action with the field theory result obtained in terms of coherent
operators. 

Before doing that, however, let us finish this section by noticing that, in principle, the spikes could be 
moving in any direction along the $S^3$ of \ads{5}:
\beq
ds^2 = -\cosh^2\!\rho\ dt^2 + d\rho^2 +\sinh^2\!\rho\ d\Omega_{[3]}^3
\eeq
When all spikes are moving  at small velocities with respect to each other this motion is again expected 
to be described by the same classical mechanics. If we go to embedding coordinates where \ads{5} is the manifold
described by:
\beq
 X_1^2+X_2^2+X_3^2+X_4^2 - X_0^2-X_{-1}^2 = -1
\eeq
 and we introduce complex coordinates $Z_1=X_1+iX_2$, $Z_2=X_3+X_4$, $W=X_0+iX_{-1}$, the solutions we described
move in $Z_1$, $W$. Since the phase of $W$ should be identified with time, the spatial sections we are interested 
in can be described as a coset $SU(1,1)/U(1)$. If we include motion in $Z_2$ this should generalize to 
$SU(2,1)/U(1)$.


\section{Gauge theory description}
\label{field theory}

\subsection{Field theory operators and coherent state description}

Now we would like to describe the operators dual to these strings in the field theory. First, for a small number 
of spikes and for $\rho_1$, namely the radial position of the tip of the spikes, approaching the boundary, 
it seems that the corresponding state in the boundary theory is that of a set of particles moving at almost the 
speed of light. One particle for each spike. Therefore, using the operator state correspondence 
the related operator, at least in the free theory, should be
\beq
\mathcal{O} = \tr \partial_+^{l_1} \Phi_1 \ldots \partial_+^{l_n} \Phi_n
\eeq
 To get a gauge invariant operator we replace ordinary derivatives by covariant 
derivatives $\partial_+\rightarrow D_+$ everywhere and recover $(\ref{ops})$ as the suggested dual operator.
 In fact, starting from the field theory, it was already suggested in \cite{Belitsky:2003ys} that
operators of this type should be described by strings with several cusps approaching the boundary. 
Here we provide actual string solutions with such property.
 The fields $\Phi_i$ can be the scalars, fermions or the field strength of \N{4} SYM theory. 
It is clear that we do not
have enough information to distinguish which particular field appears. That should be determined by other
internal degrees of freedom of the string. For the scalar they are clearly the directions of the 
sphere $S^5$. For the fermions and field strength they are presumably the fermionic degrees of freedom propagating
on the string. In any case we leave this interesting problem for the future and concentrate here on the case
where the main contribution to the anomalous dimension comes from the derivatives and properly 
identifying $\Phi$ is irrelevant. 

In that case, ignoring the contributions from the $\Phi_i$, the total number of derivatives 
determines the angular momentum of the operator
$J=\sum_{i=1}^{n} l_i$ and the conformal dimension in the free theory $\Delta=J=\sum_{i=1}^{n} n_i$. Identifying
$\Delta$ with $E$, the energy of the string we reproduce the leading result in (\ref{enres}),\ie\ $E\simeq J$, 
and are left to compute the subleading (in the large $J$ expansion) logarithmic term.

When $J$ is a multiple of $n$, an operator with $l_1=\ldots=l_n=J/n$ can be constructed. Such
configuration is invariant under cyclic permutations (for this purpose we assume that all the fields $\Phi_i$ are the same). 
Of course the trace enforces cyclic permutation symmetry for any operator but what we mean here is more than that, 
it means that all operators that are multiplied are the same. This fact and the fact that for such operator $J$ is 
a multiple of $n$ lead us to identify it, at least as a first step, with the spiky strings we described in 
previous sections. Moreover, for large $J$,
its anomalous dimension can be easily computed using, for example, the discussions 
in \cite{Beisert:2003jj,Stefanski:2004cw,Bellucci:2004qr}\footnote{
See also \cite{Belitsky:2003ys} for a relation to cusp anomalies of Wilson loops.} giving, for large $J$,
\beq
\Delta -J \simeq \frac{4\lambda}{(4\pi)^2}\, n \ln\, \frac{J}{n}
\eeq
in agreement with the subleading term in (\ref{En2}), except for the coupling constant 
dependence. In spin chain language, this state has the same occupation number in each site. The diagonal 
part of the Hamiltonian gives the desired result, the non-diagonal part being subleading for large $J$. 
 As pointed out in \cite{Belitsky:2003ys}, from the string point of view, this result depends only on the 
fact that the string configuration has $n$ cusps approaching the boundary and does not depend on the details 
of the string configuration.
 To get more information, we can try to go further and find an eigenstate or, in the case when the 
number of derivatives is large, use a coherent state approach. This was actually done by Stefanski and 
Tseytlin \cite{Stefanski:2004cw}\footnote{See also \cite{Bellucci:2004qr}for a discussion of the kinetic term.}. 
The coherent state at a given
site $i$ is a function of two variables:
\beq
\ket{\rho_i,\theta_i} = \frac{1}{\cosh\rho_i} \sum_{l_i=0}^{\infty} e^{il_i\theta_i} (\tanh\rho_i)^{l_i}\ \ket{l_i}
\eeq
where $\ket{l_i}$ is a state representing $\frac{1}{l_i!}D_+^{l_i}\Phi_i$. The average value of $l_i$ is given 
be $\langle l_i \rangle = \sinh^2\rho_i$ and we are going to consider it to be the same for all sites following
our previous discussion. Such coherent states can be used to find a classical action for the spin chain 
Hamiltonian resulting in \cite{Stefanski:2004cw}:
\beq
S =\half \sum_i \dot{\theta_i} (\cosh2\rho_i-1) 
   - \frac{2\lambda}{(4\pi)^2}\sum_{i=1}^n\ln\frac{1-\vec{n}_i.\vec{n}_{i+1}}{2}
\eeq
where $\vec{n}_i=(-\cosh2\rho_i,\sinh2\rho_i\cos\theta_i,\sinh2\rho_i\sin\theta_i)$ and the scalar product between
$\vec{n}$'s is taken with signature $(-1,1,1)$. Replacing the parameterization of $\vec{n}_i$ and considering a 
reduced situation where all $\rho_i$ are equal and large we get
\beq
S \simeq \half \sum_i \dot{\theta_i} (\cosh2\rho-1) -  \frac{2\lambda}{(4\pi)^2}
      \sum_{i=1}^n (4\rho+\ln\sin^2(\frac{\theta_{j+1}-\theta_j}{2}))
\eeq
 Notice that the number of derivatives at each site is $l = \sinh^2\rho$ 
and therefore we need $\rho\gg1$ also in the field theory. Again we see that we get perfect agreement with the 
string description, up to the dependence of the coupling constant. This is nevertheless remarkable because it means 
that we can give a string description of this operators even at small coupling. We only need to modify the radius 
of \ads{5} space. 

From the 1-loop calculation this symmetric 
operator with all $\langle l_i\rangle$ equal appears to be the one of largest conformal dimension suggesting
that the spiky strings should be unstable. This is also seen in the fact that the potential between 
spikes is attractive and therefore the situation is of unstable equilibrium.  In any case
this is not a problem for us since we propose these solutions as a dual description of certain 
operators and not as long-lived states.   

 As mentioned before, when studying a similar problem in QCD a related classical mechanics emerges. 
Some interesting solutions can be found in \cite{Korchemsky:1997yy}. The results of this paper suggest 
that those calculations should have an interpretation as the motion of semiclassical strings.

\subsection{The radial direction in the field theory?}

\FIGURE{\epsfig{file=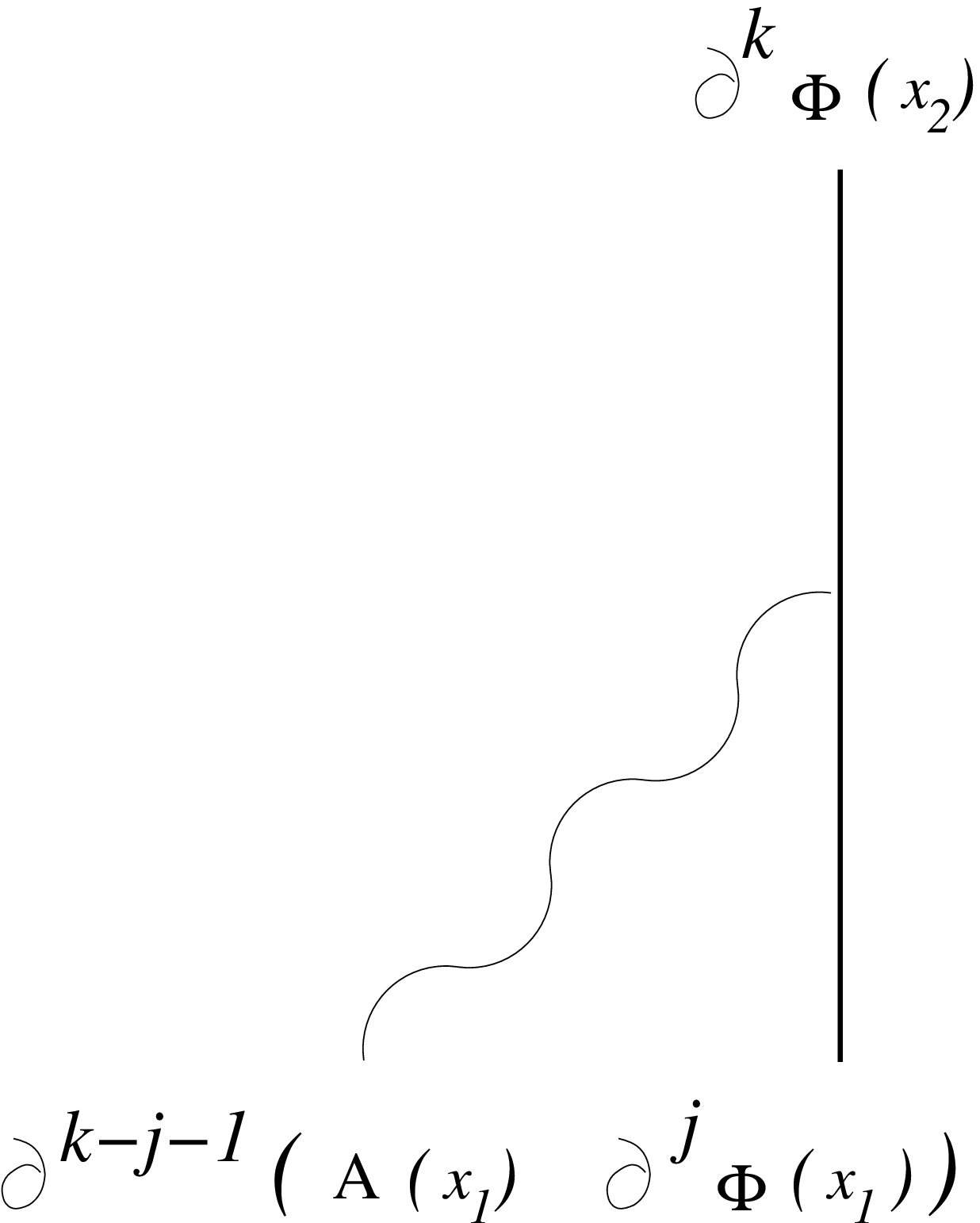, height=5cm}
\caption{Feynman diagram contributing to the anomalous dimension.}
\label{fig:diag} }

Since we have been describing a mapping between operators and strings in \ads{5} it is a natural question to
see if we can clarify in some way the meaning of the radial direction of \ads{5} from the field theory
point of view. In this subsection we give some thought to this point but unfortunately without 
arriving to any definite conclusion.

From the discussion in the previous sections a picture emerged where we can map the fields $\Phi_i$ to
cusps on the string. The radial position $\rho_1\gg1$ of the cusp is related to the number
of derivatives $l_i$ through $\sqrt{\lambda}e^{2\rho_1}\sim l_i$ simply by equating the angular
momentum of the cusp to the number of derivatives. The angular position is the
corresponding angle in the coherent state description. Although less clear,
it appears that the portion of the string hanging
between cusps is related to the gluons appearing in the covariant derivatives. Unfortunately
we cannot make this very precise but there is an interesting fact that seems to appear already
in perturbation theory. The covariant derivatives are naturally ordered because they do not commute. Notice
that this in not only because the gauge field is matrix-valued  but also because 
the ordinary derivative and the gauge field $A_\mu(x)$ do not commute. So, even at lowest order, when 
 all derivatives are ordinary except one, there is a meaning to the question 
of where the gluon is inserted. Consider the operator
\beq
\mathcal{O} = \frac{1}{k!} D_+^k\Phi = \frac{1}{k!} \partial_+^k \Phi 
 + g_{YM} \frac{1}{k!} \sum_{j=0}^{k-1} \partial_+^{k-j-1}\left(A_+\partial_+^{j}\Phi\right) +\ldots
\eeq
where we kept only the lowest term in the expansion in the Yang-Mills coupling constant $g_{YM}$.
 If we now consider a diagram such as that in fig.(\ref{fig:diag}) (and use Feynman gauge) it turns 
out that the contribution to the anomalous dimension coming from the term 
\beq
 g_{YM} \frac{1}{k!} \partial_+^{k-j-1}\left(A_+\partial_+^{j}\Phi\right)
\eeq
is proportional to $\frac{1}{j+1}+\frac{1}{j+2}$. The total contribution of this diagram to the anomalous dimension
$\delta$ is 
\beq
\delta \sim \sum_{j=0}^{k-1} \left(\frac{1}{j+1}+\frac{1}{j+2}\right) 
\eeq
which for large $k$ behaves as $2\ln k$ giving the logarithmic dependence on the angular momentum which here
is given by $k$.  If we notice that $j$ is the number of derivatives between $A_+(x)$
and the operator $\Phi$, we can loosely say that the contribution to the anomalous dimension is inversely
proportional to the distance from $A(x)$ to the operator $\Phi$ (measured by how many derivatives are between them).

If we compare with the result in the string side where we have at leading order
\beq
E-J \sim \int_{\rho_0}^{\rho_1} d\rho 
\eeq
it seems natural to identify $j\sim e^{\rho_1-\rho}$ and therefore one will be tempted to say that a gluon
corresponding to a covariant derivative ``far'' from the operator corresponds to a portion of string
deeper in \ads{5} space. Clearly more work is required to make this more precise.

\section{Conclusions}
\label{conclu}

 In this paper we studied certain single trace operators in gauge theories and argued that 
they have a dual description in terms of strings with spikes, one spike for each field
appearing in the operator. In the case of \N{4} where the dual background is known, we 
solved the equation of motion for the string and find that the motion of the spikes can be described 
by an effective classical mechanics which, in its form, agrees with a coherent state description
of the operators in the field theory generalizing previous results in \cite{GKP,Belitsky:2003ys}. Using only 
a 1-loop calculation, the dependence on the coupling constant does not match the supergravity result which in principle is not
surprising since the latter is valid for strong coupling. The main point however is that even 
the 1-loop calculations in the field theory can be interpreted as coming from a classical string moving
in a background. Computing higher loops in the field theory would change the Hamiltonian of the spin chain
but keep the generic picture intact. One exception is that at higher loops the number of operators is not 
conserved under renormalization group flow. This can be taken into account since neither is conserved the 
number of spikes. It should be interesting to see if the formation or disappearance of spikes can be matched 
to analogous processes in the field theory.

 Although the operators we have studied are not the most generic ones, it seems that the picture we advocate
here, of strings with spikes that represent local operators, should be rather generic. If instead of operators
we consider states of the theory in $S^3$, those would be localized particles moving almost at the speed of light
on the positions of the sphere closest to each spike and surrounded by a ``cloud'' corresponding to the 
the strings hanging between the tip of the spikes. This is in the case of \N{4} which is not confining. In 
a confining case it is possible that a similar picture describes localized particles connected by 
flux tubes.

\section{Acknowledgments}
I am grateful to A. Lawrence, J. Maldacena and A. Tseytlin for discussions on related matters.
I am also indebted to A. Belitsky, A. Gorsky and G. Korchemsky for various comments and 
clarifications on previous work on the subject.   
The author is supported in part by NSF under grant PHY-0331516 and by DOE under grant 
DE-FG-02-92ER40706 and a DOE Outstanding Junior Investigator Award.

\appendix

\section{Spiky strings in flat space, rigid rotation}
\label{flatNG}

 The same calculations that we did for the rotating string in AdS space can be
done in flat space and recover the solutions discussed in section \ref{flat space}.
 We use world-sheet coordinates such that 
\beq
t = \tau,\ \ \  \theta=\omega\tau+\sigma
\eeq
and use the ansatz $\rho=\rho(\sigma)$ for a string propagating in a metric
\beq
ds^2=-dt^2 + d\rho^2 + \rho^2 d\theta^2
\eeq
The equations of motion following from the Nambu-Goto action are satisfied if
\beq
\rho' = \frac{\rho\rho_1}{\rho_0}\frac{\sqrt{\rho^2-\rho_0^2}}{\sqrt{\rho_1^2-\rho^2}} 
\label{eqNGfl}
\eeq
where $\rho_1=\frac{1}{\omega}$ and $\rho_0$ is a constant of integration. 
 From the Nambu-Goto action:
\beq
S = - \int \sqrt{(1-\rho^2\omega^2)\rho'^2+\rho^2}
\eeq
we can compute the angular momentum an energy as
\beqa
J &=&\frac{1}{2\pi}\int \frac{\omega\rho^2\rho'^2}{\sqrt{(1-\rho^2\omega^2)\rho'^2+\rho^2}}\, d\sigma = 
   \frac{1}{2\pi} \int_{\rho_0}^{\rho_1} \frac{\sqrt{\rho^2-\rho_0^2}}{\sqrt{\rho_1^2-\rho^2}}\rho\,d\rho 
   = \frac{1}{8} (\rho_1^2-\rho_0^2)\\
E &=& \frac{1}{2\pi}\int \frac{\rho'^2+\rho^2}{\sqrt{(1-\rho^2\omega^2)\rho'^2+\rho^2}} d\sigma =
   \frac{1}{2\pi} \frac{1}{\rho_1} \int_{\rho_0}^{\rho_1} 
      \frac{\rho_1^2-\rho_0^2}{\sqrt{\rho_1^2-\rho^2}\sqrt{\rho^2-\rho_0^2}}\rho\,d\rho \\
   &=& \frac{1}{4}\frac{1}{\rho_1}(\rho_1^2-\rho_0^2)
\eeqa
where we used eq.(\ref{eqNGfl}). The angle difference between spike and valley can be computed as
\beq
\Delta\theta = \int d\sigma = \frac{\rho_0}{\rho_1}\int_{\rho_0}^{\rho_1} 
   \frac{\sqrt{\rho_1^2-\rho^2}}{\sqrt{\rho^2-\rho_0^2}} \frac{d\rho}{\rho} = \frac{\pi}{2} \frac{\rho_1-\rho_0}{\rho_1}
\eeq
 Now we use that $\Delta = \frac{2\pi}{n}$ where $n$ is the number of spikes that the string has. This gives
\beq
 \frac{\rho_0}{\rho_1} = 1-\frac{2}{n}
\label{rhorel}
\eeq
 which results in 
\beq
J = \frac{n-1}{n}\,\rho_1^2\, , \ \  E = 2  \frac{n-1}{n}\rho_1\ \ \ \ \ \Rightarrow \ \ \ E = 2\sqrt{\frac{n-1}{n}J} 
\eeq
which reproduces the result we obtained using conformal gauge. To see the relation more directly we can do the
following coordinate transformation on the world-sheet:
\beqa
\tau &=& x_1 + \frac{\rho_0}{\rho_1} x_2 \\
\sigma &=& -\frac{\rho_0}{\rho_1^2} x_2 + \arctan\left(\frac{\rho_0}{\rho_1}\tan \frac{x_2}{\rho_1}\right)
\eeqa
which puts the world sheet metric in diagonal form and results in 
\beqa
t &=& x_1 + \frac{\rho_0}{\rho_1} x_2 \\
\rho &=& \sqrt{\rho_1^2\cos^2\omega x_2 + \rho_0^2\sin^2\omega x_2} \\
\theta &=&  \omega x_1 +   \arctan\left(\frac{\rho_0}{\rho_1}\tan \frac{x_2}{\rho_1}\right)
\eeqa
or, in Cartesian coordinates
\beqa
 x &=& \rho \cos\theta = \half\left((\rho_1+\rho_0)\cos\omega x_+ + (\rho_1-\rho_0)\cos\omega x_-\right) \\
 y &=& \rho \sin\theta = \half\left((\rho_1+\rho_0)\sin\omega x_+ + (\rho_1-\rho_0)\sin\omega x_-\right) \\
\eeqa
where $x_\pm=x_1\pm x_2$.
 Finally we do a last coordinate transformation given by
\beqa
 \tilde{\tau} &=& x_1 + \frac{\rho_0}{\rho_1} x_2 \\
 \tilde{\sigma} &=& \frac{\rho_0}{\rho_1} x_1 + x_2 
\eeqa
which results in 
\beqa
x &=& \half\left\{(\rho_1+\rho_0)\cos(\frac{\tilde{\tau}+\tilde{\sigma}}{\rho_1+\rho_0}) +
                 (\rho_1-\rho_0)\cos(\frac{\tilde{\tau}-\tilde{\sigma}}{\rho_1-\rho_0})\right\} \\
y &=& \half\left\{(\rho_1+\rho_0)\sin(\frac{\tilde{\tau}+\tilde{\sigma}}{\rho_1+\rho_0}) +
                 (\rho_1-\rho_0)\sin(\frac{\tilde{\tau}-\tilde{\sigma}}{\rho_1-\rho_0})\right\} \\
t &=& \tilde{\tau}
\eeqa
which is the same as (\ref{flatsol}) after identifying 
\beq
\tilde{\tau} = (\rho_1+\rho_0)\tau, \ \ \ \tilde{\sigma}=-(\rho_1+\rho_0)\sigma, \ \ \ \lambda=\half(\rho_1-\rho_0)
\eeq
and using, from (\ref{rhorel}), that
\beq
\frac{\rho_1+\rho_0}{\rho_1-\rho_0}=n-1\ .
\eeq


\begin{thebibliography}{99}        


\bibitem{largeN}
G.'t Hooft, Nucl. Phys. {\bf B72} (1974) 461, 
G.'t Hooft, Nucl. Phys. {\bf B75} (1974) 461. 

\bibitem{malda}
J.~Maldacena,
``The large $N$ limit of superconformal field theories and supergravity,''
Adv.\ Theor.\ Math.\ Phys.\  {\bf 2}, 231 (1998)
[Int.\ J.\ Theor.\ Phys.\  {\bf 38}, 1113 (1998)],
{\tt hep-th/9711200}, \\
S.~S.~Gubser, I.~R.~Klebanov and A.~M.~Polyakov,
``Gauge theory correlators from non-critical string theory,''
Phys.\ Lett.\ B {\bf 428}, 105 (1998)
[arXiv:hep-th/9802109], \\
E.~Witten,
``Anti-de Sitter space and holography,''
Adv.\ Theor.\ Math.\ Phys.\  {\bf 2}, 253 (1998)
[arXiv:hep-th/9802150], \\
O.~Aharony, S.~S.~Gubser, J.~M.~Maldacena, H.~Ooguri and Y.~Oz,
``Large N field theories, string theory and gravity,''
Phys.\ Rept.\  {\bf 323}, 183 (2000)
[arXiv:hep-th/9905111].


\bibitem{Berenstein:2002jq}
D.~Berenstein, J.~M.~Maldacena and H.~Nastase,
``Strings in flat space and pp waves from N = 4 super Yang Mills,''
JHEP {\bf 0204}, 013 (2002)
[arXiv:hep-th/0202021].

\bibitem{GKP}
S.~S.~Gubser, I.~R.~Klebanov and A.~M.~Polyakov,
``A semi-classical limit of the gauge/string correspondence,''
Nucl.\ Phys.\ B {\bf 636}, 99 (2002)
[arXiv:hep-th/0204051].

\bibitem{Tseytlin:2003ii}
A.~A.~Tseytlin,
``Spinning strings and AdS/CFT duality,''
arXiv:hep-th/0311139.

\bibitem{Tseytlin:2004xa}
A.~A.~Tseytlin,
``Semiclassical strings and AdS/CFT,''
arXiv:hep-th/0409296.

\bibitem{Tseytlin:2004cj}
A.~A.~Tseytlin,
``Semiclassical strings in AdS(5) x S**5 and scalar operators in N = 4 SYM
theory,''
arXiv:hep-th/0407218.

\bibitem{Minahan:2002ve}
J.~A.~Minahan and K.~Zarembo,
``The Bethe-ansatz for N = 4 super Yang-Mills,''
JHEP {\bf 0303} (2003) 013
[arXiv:hep-th/0212208].

\bibitem{Braun:1998id}
V.~M.~Braun, S.~E.~Derkachov and A.~N.~Manashov,
``Integrability of three-particle evolution equations in {QCD},''
Phys.\ Rev.\ Lett.\  {\bf 81}, 2020 (1998)
[arXiv:hep-ph/9805225], \\
V.~M.~Braun, S.~E.~Derkachov, G.~P.~Korchemsky and A.~N.~Manashov,
``Baryon distribution amplitudes in {QCD},''
Nucl.\ Phys.\ B {\bf 553}, 355 (1999)
[arXiv:hep-ph/9902375],\\
A.~V.~Belitsky,
``Integrability and WKB solution of twist-three evolution equations,''
Nucl.\ Phys.\ B {\bf 558}, 259 (1999)
[arXiv:hep-ph/9903512], \\
A.~V.~Belitsky,
``Fine structure of spectrum of twist-three operators in {QCD},''
Phys.\ Lett.\ B {\bf 453}, 59 (1999)
[arXiv:hep-ph/9902361].

\bibitem{Beisert:2003ea}
N.~Beisert, S.~Frolov, M.~Staudacher and A.~A.~Tseytlin,
``Precision spectroscopy of AdS/CFT,''
JHEP {\bf 0310}, 037 (2003)
[arXiv:hep-th/0308117].

\bibitem{Beisert:2003xu}
N.~Beisert, J.~A.~Minahan, M.~Staudacher and K.~Zarembo,
``Stringing spins and spinning strings,''
JHEP {\bf 0309}, 010 (2003)
[arXiv:hep-th/0306139].

\bibitem{Beisert:2003tq}
N.~Beisert, C.~Kristjansen and M.~Staudacher,
``The dilatation operator of N = 4 super Yang-Mills theory,''
Nucl.\ Phys.\ B {\bf 664}, 131 (2003)
[arXiv:hep-th/0303060].

\bibitem{Beisert:2003jj}
N.~Beisert,
``The complete one-loop dilatation operator of N = 4 super Yang-Mills
theory,''
Nucl.\ Phys.\ B {\bf 676}, 3 (2004)
[arXiv:hep-th/0307015].

\bibitem{Frolov:2003xy}
S.~Frolov and A.~A.~Tseytlin,
``Rotating string solutions: AdS/CFT duality in non-supersymmetric  sectors,''
Phys.\ Lett.\ B {\bf 570}, 96 (2003)
[arXiv:hep-th/0306143].

\bibitem{Arutyunov:2003uj}
G.~Arutyunov, S.~Frolov, J.~Russo and A.~A.~Tseytlin,
``Spinning strings in AdS(5) x S**5 and integrable systems,''
Nucl.\ Phys.\ B {\bf 671}, 3 (2003)
[arXiv:hep-th/0307191].

\bibitem{Arutyunov:2003rg}
G.~Arutyunov and M.~Staudacher,
``Matching higher conserved charges for strings and spins,''
JHEP {\bf 0403}, 004 (2004)
[arXiv:hep-th/0310182].

\bibitem{Kruczenski:2003gt}
M.~Kruczenski,
``Spin chains and string theory,'', 
Phy. Rev. Lett {\bf 93}, 161602 (2004),
[arXiv:hep-th/0311203].

\bibitem{Dimov:2004qv}
H.~Dimov and R.~C.~Rashkov,
``A note on spin chain / string duality,''
arXiv:hep-th/0403121.

\bibitem{Hernandez:2004uw}
R.~Hernandez and E.~Lopez,
``The SU(3) spin chain sigma model and string theory,''
JHEP {\bf 0404}, 052 (2004)
[arXiv:hep-th/0403139].

\bibitem{Ryang:2004tq}
S.~Ryang,
``Folded three-spin string solutions in AdS(5) x S**5,''
JHEP {\bf 0404}, 053 (2004)
[arXiv:hep-th/0403180].

\bibitem{Dimov:2004xi}
H.~Dimov and R.~C.~Rashkov,
``Generalized pulsating strings,''
JHEP {\bf 0405}, 068 (2004)
[arXiv:hep-th/0404012].

\bibitem{Stefanski:2004cw}
B.~.~J.~Stefanski and A.~A.~Tseytlin,
``Large spin limits of AdS/CFT and generalized Landau-Lifshitz equations,''
JHEP {\bf 0405}, 042 (2004)
[arXiv:hep-th/0404133].

\bibitem{Kruczenski:2004cn}
M.~Kruczenski and A.~A.~Tseytlin,
``Semiclassical relativistic strings in S**5 and long coherent operators in N
= 4 SYM theory,''
JHEP {\bf 0409}, 038 (2004)
[arXiv:hep-th/0406189].

\bibitem{Ideguchi:2004wm}
K.~Ideguchi,
``Semiclassical strings on AdS(5) x S**5/Z(M) and operators in orbifold field
theories,''
JHEP {\bf 0409}, 008 (2004)
[arXiv:hep-th/0408014].

\bibitem{Ryang:2004pu}
S.~Ryang,
``Circular and folded multi-spin strings in spin chain sigma models,''
arXiv:hep-th/0409217.

\bibitem{Hernandez:2004kr}
R.~Hernandez and E.~Lopez,
``Spin chain sigma models with fermions,''
arXiv:hep-th/0410022.

\bibitem{Bellucci:2004qr}
S.~Bellucci, P.~Y.~Casteill, J.~F.~Morales and C.~Sochichiu,
``sl(2) spin chain and spinning strings on AdS(5) x S**5,''
arXiv:hep-th/0409086.

\bibitem{Susaki:2004tg}
Y.~Susaki, Y.~Takayama and K.~Yoshida,
``Open semiclassical strings and long defect operators in AdS/dCFT
correspondence,''
arXiv:hep-th/0410139.

\bibitem{Kruczenski:2004kw}
M.~Kruczenski, A.~V.~Ryzhov and A.~A.~Tseytlin,
``Large spin limit of AdS(5) x S**5 string theory and low energy expansion of
ferromagnetic spin chains,''
Nucl.\ Phys.\ B {\bf 692}, 3 (2004)
[arXiv:hep-th/0403120].

\bibitem{Kazakov:2004qf}
V.~A.~Kazakov, A.~Marshakov, J.~A.~Minahan and K.~Zarembo,
``Classical / quantum integrability in AdS/CFT,''
JHEP {\bf 0405}, 024 (2004)
[arXiv:hep-th/0402207].

\bibitem{Kazakov:2004nh}
V.~A.~Kazakov and K.~Zarembo,
``Classical/quantum integrability in non-compact sector of AdS/CFT,''
arXiv:hep-th/0410105.

\bibitem{Ferretti:2004ba}
G.~Ferretti, R.~Heise and K.~Zarembo,
``New integrable structures in large-N QCD,''
arXiv:hep-th/0404187.

\bibitem{Belitsky:2004cz}
A.~V.~Belitsky, V.~M.~Braun, A.~S.~Gorsky and G.~P.~Korchemsky,
``Integrability in QCD and beyond,''
arXiv:hep-th/0407232.

\bibitem{Mikhailov:2004qf}
A.~Mikhailov,
``Slow evolution of nearly-degenerate extremal surfaces,''
arXiv:hep-th/0402067.

\bibitem{Mikhailov:2004xw}
A.~Mikhailov,
``Supersymmetric null-surfaces,''
JHEP {\bf 0409}, 068 (2004)
[arXiv:hep-th/0404173].

\bibitem{Mikhailov:2004au}
A.~Mikhailov,
``Notes on fast moving strings,''
arXiv:hep-th/0409040.

\bibitem{Gorsky:2003nq}
A.~Gorsky,
``Spin chains and gauge / string duality,''
arXiv:hep-th/0308182.

\bibitem{Mateos:2003de}
D.~Mateos, T.~Mateos and P.~K.~Townsend,
``Supersymmetry of tensionless rotating strings in AdS(5) x S**5, and nearly-BPS operators,''
arXiv:hep-th/0309114.

\bibitem{Kruczenski:2002fb}
M.~Kruczenski,
``A note on twist two operators in N = 4 SYM and Wilson loops in Minkowski
signature,''
JHEP {\bf 0212}, 024 (2002)
[arXiv:hep-th/0210115].

\bibitem{Makeenko:2002qe}
Y.~Makeenko,
``Light-cone Wilson loops and the string / gauge correspondence,''
JHEP {\bf 0301}, 007 (2003)
[arXiv:hep-th/0210256].

\bibitem{Korchemsky}
G.~P.~Korchemsky,
``Asymptotics Of The Altarelli-Parisi-Lipatov Evolution Kernels Of Parton Distributions,''
Mod.\ Phys.\ Lett.\ A {\bf 4}, 1257 (1989); \\
G.~P.~Korchemsky and G.~Marchesini,
``Structure function for large x and renormalization of Wilson loop,''
Nucl.\ Phys.\ B {\bf 406}, 225 (1993)
[arXiv:hep-ph/9210281].


\bibitem{DIS}
H. Georgi and H.D. Politzer, Phys. Rev. {\bf D9} (1974) 416, \\
D.J. Gross and F. Wilczek, Phys. Rev. {\bf D9}  (1974) 920. 

\bibitem{Belitsky:2003ys}
A.~V.~Belitsky, A.~S.~Gorsky and G.~P.~Korchemsky,
``Gauge / string duality for QCD conformal operators,''
Nucl.\ Phys.\ B {\bf 667}, 3 (2003)
[arXiv:hep-th/0304028].


\bibitem{Grad}
I.S. Gradshteyn, I.M. Ryzhik, ``Table of Integrals Series and Products'', Sixth edition,
Academic Press (2000), San Diego, CA, USA, London, UK.


\bibitem{Korchemsky:1997yy}
  G.~P.~Korchemsky and I.~M.~Krichever,
  ``Solitons in high-energy {QCD},''
  Nucl.\ Phys.\ B {\bf 505}, 387 (1997)
  [arXiv:hep-th/9704079].

\bibitem{flatc}
  C.~J.~Burden,
  ``Gravitational Radiation From A Particular Class Of Cosmic Strings,''
  Phys.\ Lett.\ B {\bf 164}, 277 (1985), \\
  C.~J.~Burden and L.~J.~Tassie,
  ``Additional Rigidly Rotating Solutions In The String Model Of Hadrons,''
  Austral.\ J.\ Phys.\  {\bf 37}, 1 (1984).



\end{thebibliography}
\end{document}